# Analytical model of misinformation of a social network node

Yuri Monakhov[1], Maria. Medvednikova[1], Konstantin Abramov[1], Natalia Kostina[1],
Roman Malyshev[1], Makarov Oleg[1], Irina Semenova[1]

[1] *Department of Informatics and Information security, Vladimir State University, Vladimir 600000, Russia*

This paper presents the research of the influence of cognitive, behavioral, representational factors on the susceptibility of the participants in social networks to misinformation, as well as on the activity of the nodes in this regard. The importance of this research consists of method of blocking the propaganda. This is very important because when people involuntarily acquire information some of them experience an undesired change in their social attitude. Such phenomena typically lead towards the information warfare. A model was developed during this research for calculating the level of misinformation of the social network participant (network node) based on the model of iterative learning process.

## I. INTRODUCTION

The purpose of this paper is researching of the influence of cognitive, behavioral, representational factors on the susceptibility of the participants in social networks to misinformation, as well as on the activity of the nodes in this regard. The importance of this research lies in a method of blocking the propaganda. It is very important because when people involuntarily acquire information some of them experience an undesired change in their social attitude. Such phenomena typically lead towards the information warfare. A model for calculating the level of misinformation of the social network participant (network node) based on the model of iterative learning process was developed during this research.

## II. ANALYTICAL MODEL OF MISINFORMATION OF A SOCIAL NETWORK NODE

There is noticeable development in researches of misinformation in social networks nowadays. Universal models for the study of this process have not been created. In this research the authors propose an analytical model of misinformation in social network of node that considers social and psychological characteristics of the person. In the course the descriptive model of R. Bush and F. Mosteller [1] was chosen.

In the descriptive models, assumptions are often based on intuition and appeal to common sense.

Model is described by equation (1):

$$\lambda_i(t) = \xi \cdot \frac{\alpha \cdot \lambda^0}{\lambda^0 + (\alpha - \lambda^0) \cdot \exp(-\alpha \cdot <Tr_{ij}> \cdot <\lambda'_j> \cdot B \cdot t)}, \quad (1)$$

$$\lambda_i(t) \in [0,1]$$

$\xi$ – coefficient reflecting if the user stays in the social network at a time. It is set to "0" or "1" and indicates whether the node is online.;

$\alpha$ – misinformation level threshold. If the node exceeds this level it starts to spread misinformation;

$\lambda^0$ - initial value of misinformation. It indicates original knowledge about the subject of propaganda;

$Tr_{ij}$ – confidence coefficient of j-th source. It indicates the degree of confidence in the source of propaganda. $Tr_{ij}$ is calculated by the formula (2):

$$Tr_{ij} = \frac{2 \cdot Mut_{ij}}{k_i + k_j} \quad (2)$$

$Mut_{ij}$ - the number of mutual nodes.

$k_i$ - the number of links on the i-th node

degree. $k_j$ - the number of links on the j-node.

$\lambda'_j$ – misinformation of j-th source;

$B$ – integral coefficient of the behavioral, cognitive and representational characteristics of node. Coefficient B indicates influence of different formulations of propaganda on the learning for the individual user. It is calculated as a weighted sum of appropriate social factors, which in turn are devised on the basis of the sociological survey.

$$B = \begin{pmatrix} V_f \\ V_p \\ V_\varphi \\ V_\tau \\ V_k \\ V_s \\ V_v \end{pmatrix} \cdot \begin{pmatrix} f \\ p \\ \varphi \\ \tau \\ k \\ s \\ v \end{pmatrix} \quad (3)$$

$V_f, V_p, V_\varphi, V_\tau, V_k, V_s, V_v$ are the weights of socio-psychological factors. $\sum_i V_i = 1$.

f, p, φ, τ, k, s, v are socio-psychological parameters.

f (f ∈ [0,1]) is a parameter of variability of misinformation. It refers to the influence of different formulations of propaganda on learning of node.

p (p ∈ [0,1]) is a parameter of targeting of misinformation. It refers to the influence of the correlation of misinformation profile data on the learning.

φ (φ ∈ [0,1]) is a parameter of the psychological vulnerability to misinformation. It refers to the influence of psychological characteristics on the perception of misinformation.

τ (τ ∈ [0,1]) is a parameter of the level of knowledge about information flow. It refers to the influence of the level of knowledge about the specifics of information flow.

k (k ∈ [0,1]) is a parameter of the competence of node in the subject of propaganda. It refers to the influence of the competence in some subject on learning of node.

s (s ∈ [0,1]) is a parameter of perception which depends on the form of information representation. It gives an estimate of influence of representation form.

v (v ∈ [0,1]) is a parameter of perception which depends on the amount of information. It refers to the influence of amount of the represented information on learning results.

Values of social and psychological parameters were determined by a survey each question of which is estimated by numeric parameters. The test questions are compiled with the sociologist.

During the sociological data processing we encounter some difficulties. The problem was that one question affects multiple social and psychological parameters. In order to obtain the distribution of the values of the parameters, a special algorithm was produced:

There are n - the number of questions, $\bar\tau_i$ - the parameter vectors, $w_i$ - weight issues, $\bar p_i$ - statistics vectors of the i-th question τ- the value of the parameter. And we have to find the fraction of P (τ) node parameter τ.

1. Sort vector $\bar\tau_i$ and $\bar p_i$ in ascending
2. $\tau_i$: $\tau_i = w_i \cdot \tau_i$.
3. $P_i(\tau^*)$ is calculated by the formula:

$$P_i(\tau^*) = \begin{bmatrix} 0, \tau_{\min}(\tau_i), \\ k_{ij} \cdot \tau^* + b_{ij}, \tau_i < \tau^* < \tau_j, k_{ij} = \dfrac{p_i - p_j}{\tau_i - \tau_j}; b_{ij} = p_j - k_{ij}\tau_j \\ 1, \tau^* > \tau_{\max}(\tau_i) \end{bmatrix}$$

4. $P(\tau)$ is calculated by the formula:
$$P(\tau) = <P_i(\tau_i^*)>$$

For example: How many respondents have the $\tau = 0{,}15$?

The survey question number one:
$w = 0{,}4$

$$\tau_1 = \begin{pmatrix} 0{,}1 \\ 0{,}2 \\ 0{,}5 \\ 0{,}7 \\ 0{,}9 \end{pmatrix}, \quad P_1 = \begin{pmatrix} 0{,}05 \\ 0{,}2 \\ 0{,}13 \\ 0{,}12 \\ 0{,}5 \end{pmatrix}.$$

The survey question number two:

$w = 0{,}6$

$$\tau_2 = \begin{pmatrix} 0{,}1 \\ 0{,}3 \\ 0{,}7 \\ 0{,}9 \end{pmatrix}, \quad P_1 = \begin{pmatrix} 0{,}3 \\ 0{,}15 \\ 0{,}15 \\ 0{,}4 \end{pmatrix}.$$

Using step number two, we can calculate $\tau^*_1$:

The survey question number one:
$w = 0{,}4$

$$\tau_1 = \begin{pmatrix} 0{,}1 \\ 0{,}2 \\ 0{,}5 \\ 0{,}7 \\ 0{,}9 \end{pmatrix}; \quad \tau^*_1 = \begin{pmatrix} 0{,}04 \\ 0{,}08 \\ 0{,}2 \\ 0{,}28 \\ 0{,}36 \end{pmatrix}.$$

The survey question number two:
$w = 0{,}6$

$$\tau_2 = \begin{pmatrix} 0{,}1 \\ 0{,}3 \\ 0{,}7 \\ 0{,}9 \end{pmatrix}; \quad \tau^*_2 = \begin{pmatrix} 0{,}06 \\ 0{,}18 \\ 0{,}42 \\ 0{,}54 \end{pmatrix}.$$

$\tau^*_1 = w_1 \cdot \tau = 0{,}15 \cdot 0{,}4 = 0{,}06$.

$0{,}04 < 0{,}06 < 0{,}08 \Rightarrow \tau_i = 0{,}04; \tau_j = 0{,}08 \Rightarrow$
$P_i = 0{,}05; P_j = 0{,}2$

$k_{ij} = (0{,}05 - 0{,}2)/(0{,}04 - 0{,}08) = 3{,}75$
$b_{ij} = 0{,}2 - 3{,}75 \cdot 0{,}08 = -0{,}1$
$P_1(\tau^*_1) = 3{,}75 \cdot 0{,}06 - 0{,}1 = 0{,}125$
$P_2(\tau^*_2) = -1{,}25 \cdot 0{,}09 + 0{,}375 = 0{,}263$.

$P(\tau) = <P_i(\tau^*_i)>$.
$P(\tau) = (0{,}125 + 0{,}263)/2 \approx 19\%$.

This algorithm was later implemented in programs written in Visual Basic in Microsoft Excel 2010.

Result of the program for parameter «τ» is presented at the picture.

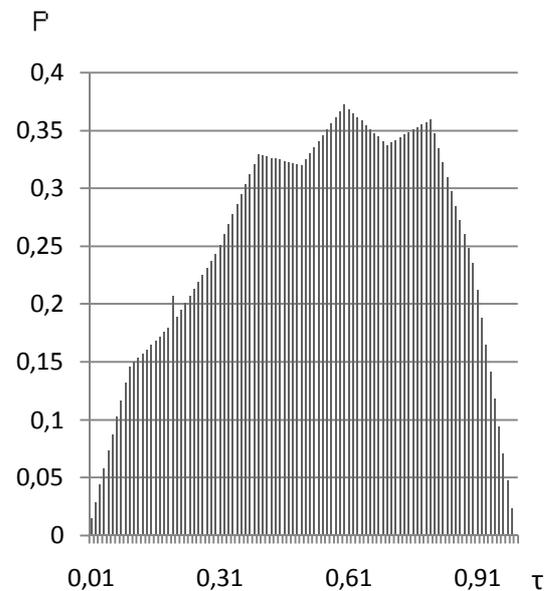

## CONCLUSIONS

In this work the number of tasks was executed: A model of a social network site misinformation was described, behavioral, cognitive, and representational parameters were identified and described, way to weigh links in the social graph was defined, the algorithm data obtained during the survey required for the calculation of social and Psychological parameters was developed; a program that builds graphs of relations of these parameters and statistics was written, example of the calculation of misinforming the i-node was described. In the future, the results of this research will help in improving the functional stability of the network.

## REFERENCES

[1] Robert R. Bush and Frederick Mosteller A Stochastic Model with Applications to Learning// Ann. Math. Statist.,Vol. 24, N 4 (1953), P. 559-585/ Harvard University.